\documentclass[twocolumn,aps,prb,floatfix,superscriptaddress,showpacs]{revtex4}
\usepackage{graphicx}
\usepackage{dcolumn}
\usepackage{bm}
\usepackage{amsfonts}
\usepackage{amssymb}
\usepackage{amsmath}

\usepackage{epsfig}

\newcommand{\cmtx}{Hg$_{\scriptstyle 1-x}$Cd$_{\scriptstyle x}$Te\,}

\newcommand{\trieq}{\mathrel{\widehat{=}}}
\DeclareMathAlphabet{\mathitb}{OT1}{cmr}{bx}{sl}

\begin{document}
\pagenumbering{arabic}
  \title{Tunable quantum spin Hall effect in double quantum wells}
  \author{Paolo~Michetti}
  \affiliation{Institute of Theoretical Physics and Astrophysics, University of W\"urzburg, D-97074 W\"urzburg, Germany}
  \author{Jan~C.~Budich}
   \affiliation{Institute of Theoretical Physics and Astrophysics, University of W\"urzburg, D-97074 W\"urzburg, Germany}
   \author{Elena~G.~Novik}
   \affiliation{Physical Institute, University of W\"urzburg, D-97074 W\"urzburg, Germany}
  \author{Patrik~Recher}
  \affiliation{Institute of Theoretical Physics and Astrophysics, University of W\"urzburg, D-97074 W\"urzburg, Germany}
\affiliation{Institute for Mathematical Physics, TU Braunschweig, 38106 Braunschweig, Germany}
  \pacs{73.20.-r, 73.43.-f, 73.21.Fg, 73.61.-r}
  
  \date{\today}
  
  \begin{abstract}
    The field of topological insulators (TIs) is rapidly growing. 
    Concerning possible applications, the search for materials with an easily controllable TI phase is a key issue.    
    The quantum spin Hall effect, characterized by a single pair of helical edge modes protected by time-reversal symmetry, 
    has been demonstrated in HgTe-based quantum wells (QWs) with an inverted bandgap.
    We analyze the topological properties of a generically coupled HgTe-based double QW (DQW) and show how in such a system a TI phase can be 
    driven by an inter-layer bias voltage, even when the individual layers are non-inverted. 
    We argue, that this system allows for similar (layer-)pseudospin based physics as in bilayer graphene but with the crucial absence of a valley degeneracy.
  \end{abstract}
  
 \maketitle

  Since the understanding of the topological nature of the quantum Hall effect \cite{Laughlin1981,TKNN1982}, 
  topological phases have become one of the most active research fields in condensed matter physics.
  More recently, a new topological phase preserving time-reversal symmetry (TRS), the quantum spin Hall (QSH) phase \cite{kane2005a,kane2005b} has been discovered.
  The QSH phase has been theoretically predicted~\cite{bernevig2006} and experimentally realized in 2D HgTe/CdTe QWs~\cite{konig2007}. 
  The crucial ingredient of this narrow gap semiconductor material is strong spin-orbit coupling, which  
  determines the inverted band structure of HgTe.
  The experimentally accessible parameter tuning the band structure from normal (CdTe-like) to inverted is the thickness of the HgTe QW.   
  One year later, 3D TIs supporting chiral fermions as surface states have been proposed 
  and observed \cite{fuinversion2007,hsieh2008,zhang2009,hsieh2009,xia2009,chen2009}.
  These two phenomena are examples of the general concept of a TI which is a TRS preserving system with a bulk insulating 
  gap which features topologically protected edge states due to the Atiyah-Singer index theorem which is in this context 
  referred to as the bulk boundary correspondence.
  TI phases are characterized by a $\mathbb Z_2$ topological invariant~\cite{kane2005a,Moore2007}.
 
  In this work, we extend the Bernevig-Hughes-Zhang (BHZ) model~\cite{bernevig2006} to account for a double QW (DQW) of HgTe (see Fig.~1).
  We analyze how the topological features of the DQW depend on a generic tunneling Hamiltonian effectively connecting the two wells
  and on the applied voltage $V$ between them. 
  In particular, we show that a QSH phase can be driven by $V$ also when the two QWs are individually trivial.
  We also derive a reduced 2-band model that well captures the topological features of the system. 
  We give analytical expressions of the effective parameters that map the reduced model to the BHZ-model of a single HgTe QW [in Eq.~(10)]
  as a function of the tunneling matrix elements and voltage $V$.
  We show that a non-trivial $\mathbb Z_2$ topological invariant is accompanied with the appearance of a single pair of 
  helical edge states for which we calculate the energy dispersion and spinors. 
  In Fig.~\ref{fig:C1}(a) we show the phase diagram of the model as a function of tunneling amplitude and applied voltage.

  \begin{figure}[tbp]
    \centering	
    \includegraphics[width=5.3cm]{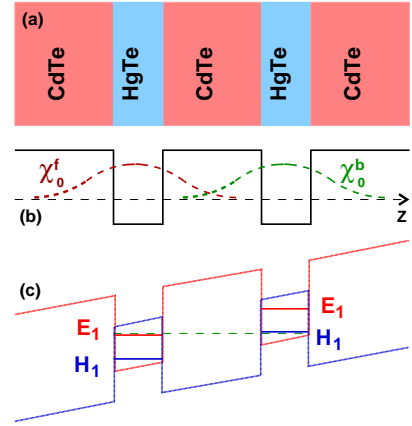}
    \caption{(Color online)
          (a) Schematic representation of the HgTe/CdTe DQW made by a front and back well 
      with a finite overlap of their individual envelope functions $\chi_0^f(z)$ and $\chi_0^b(z)$ as shown in (b).
      (c) Process of band inversion in a DQW caused by a potential bias. 
    }
    \label{fig:device}
  \end{figure}

  \section{ Double quantum well model}
  The system under investigation consists of a DQW of HgTe as shown in Fig.~\ref{fig:device}, 
  which can be thought of as a 2D TI bilayer.
  The system is contacted to gates that control the potential bias $V$ between the front (f) and back (b) QWs.
  %
  The spectrum of a single HgTe-based QW near the $\Gamma$-point  is effectively described
  by the BHZ model~\cite{bernevig2006} 
  \begin{eqnarray}
    H_0&=& \left(\begin{array}{cc}
      h(\vec{k}) &0\\
      0 & h^*(-\vec{k})
      \end{array}\right)\nonumber\\
    h(k)&=&  \vec{d}\cdot\vec{\sigma}\label{eq:H0}\\
    \vec{d}&=& \left( C-D k^2, A k_x , -A k_y   , M-B k^2 \right)\nonumber,
  \end{eqnarray}
  where $\vec{\sigma}$ are the Pauli matrices~\cite{pauli} associated with the band-pseudospin 
  degree of freedom (band $E_1$ or $H_1$).
  $H_0$ is represented in the basis $\big\{|E_1+\rangle$, $|H_1+\rangle$, $|E_1-\rangle$, $|H_1-\rangle\big\}$, 
  where the $E_1$  states ($J_z=\pm1/2$) are a mixing of the s-like $\Gamma_6$ band with the $\Gamma_8$ light-hole 
  band, while $H_1$ ($J_z=\pm 3/2$) is basically the $\Gamma_8$ heavy-hole band~\cite{symmetry}. 
  In the following we will use the parameter values $A=375$ meV nm, $B=-1.120$ eV ${\rm nm}^2$ 
  and $D=-730$ meV ${\rm nm}^2$, estimated by a comparison with the $8 \times 8$ Kane Hamiltonian~\cite{novik2005}, 
  and assume $C=0$ without loss of generality. 
  The Dirac rest mass $M$ depends on the QW thickness and $M<0$ corresponds to the inverted (QSH) regime whereas  $M>0$ corresponds to the normal regime.
  In first approximation, $H_0$ is block diagonal in the Kramer's partner or spin degree of freedom~\cite{bernevig2006} (with $\vec{s}$ the vector of Pauli matrices).
  Because we consider only systems with TRS, we restrict ourselves to the block $h(\vec{k})$, 
  from which the results can be extended to the other one by applying the time reversal operator $\hat{T}=i s_y  \hat{K}$.

  The Hamiltonian $h(\vec{k})$ in Eq.~\ref{eq:H0} describes the in-plane electronic motion inside a QW layer.
  Along the confinement direction of the QW ($Z$), electrons are described by the envelope function $\chi_0(z)$ (integrated out in the BHZ model).
  When the two QWs of a DQW are sufficiently separated, an electron will be localized either on the f or on the b layer, 
  with envelope function $\chi_0^f(z)$ and $\chi_0^b(z)$, resp.
  When the two layers are placed close to each other, $\chi_0^f(z)$ and $\chi_0^b(z)$ acquire a finite  
  overlap, accounted for by a tunneling Hamiltonian $H_T$.
  This description should take into account that $\chi_0(z)$ is a spinor 
  with components on the $E_1$ and $H_1$ bands.
  To first order in $k$,  the tunneling Hamiltonian has the form
  \begin{eqnarray} 
    H_T &=&   +\frac{\mathfrak{R}(\vec{\Delta})\cdot\vec{\sigma}}{2} \mathcal{P}_x -\frac{\mathfrak{I}(\vec{\Delta})\cdot\vec{\sigma}}{2} \mathcal{P}_y
    \label{eq:tunneling0}\\
    \vec{\Delta} &=& \left( \Delta_{0},\alpha k_x,  -\alpha  k_y,  \Delta_{z} \right), \nonumber 
 \end{eqnarray}
  where $\mathcal{P}_i$ are Pauli matrices associated with the layer projection [layer pseudospin (LPS) ${\vec{\cal P}}$].
  $H_T$ generates bonding/antibonding states with energy splitting $\Delta_{E1}=\Delta_0+\Delta_z$ ($\Delta_{H1}=\Delta_0-\Delta_z$)\cite{note2} 
  of the $E1$ ($H1$) band.
  The lowest order off-diagonal term is $\alpha k_\pm$, due to the axial symmetry and the $J_z$ character of $E1$ and $H1$ bands~\cite{symmetry}. 
  In the Appendix A, we present a realistic estimate of HgTe DQW tunneling parameters.

  The DQW Hamiltonian for a single Kramer's block is therefore 
  \begin{equation} 
    H = \vec{d}\cdot\vec{\sigma}\mathcal{P}_0 +\frac{\mathfrak{R}(\vec{\Delta})\cdot\vec{\sigma}}{2} \mathcal{P}_x -\frac{\mathfrak{I}(\vec{\Delta})\cdot\vec{\sigma}}{2} \mathcal{P}_y + \frac{1}{2}(\delta\vec{d}\cdot\vec{\sigma})\mathcal{P}_z,
    \label{eq:Htot}
  \end{equation}
  where $\delta \vec{d}$ is due to the possible variation of the parameters of $h(k)$ between the f and the b layer.
  In particular, we consider $\delta \vec{d} = (V,0,0,0)$, with $V$ the interlayer bias.

  \section{Band structure and topology} 
  The topological properties of a fully gapped 2D TRS preserving system are described by the $\mathbb Z_2$-invariant $\nu$~ \cite{fuinversion2007}.
  When the model Hamiltonian is block diagonal with respect to the Kramers partner spin $\vec s$,
  the system can be thought of, from a topological point of view, as two copies of an anomalous quantum Hall effect, related by time reversal \cite{Qi2006}.
  Each of the blocks is topologically characterized by its first Chern number $\mathcal C_{\uparrow}$~and $\mathcal C_{\downarrow}$~resp., 
  which we call the Kramers Chern numbers (KCNs). 
  TRS immediately implies that the two KCNs obey a zero sum rule~\cite{Avron1988}, meaning $\mathcal C_{\uparrow}=-\mathcal C_{\downarrow}$. 
  The $\mathbb Z_2$~invariant is then given by \cite{Qi2006}
  \begin{equation}
    \nu = \frac{\mathcal C_\uparrow-\mathcal C_\downarrow}{2} (\text{mod } 2)= \mathcal C_\uparrow (\text{mod } 2),
    \label{eq:nu}
  \end{equation}
  with the KCN defined as
  \begin{equation}
    \mathcal C = \frac{i}{2\pi}\int_{T^2}\mathcal F, 
    \label{eq:KCNDef}
  \end{equation}
  and where the Berry curvature is given by
  \begin{equation}
    \mathcal F(k) = \sum_{\alpha \text{ occ}}\left(d\langle u_k^\alpha \rvert\right)\wedge\left(d\lvert u_k^\alpha\rangle\right),
  \end{equation}
  where $\wedge$ stands for the exterior product.
  The sum is over the occupied bands and $\lvert u_k^\alpha\rangle$~are Bloch states of the Kramers block Hamiltonian. 
  In order to make the Chern number well defined in local models, like Eq.~\ref{eq:Htot}, 
  a lattice regularization has to be included to compactify the k-space. 
  In our case, since the curvature decays rapidly away from the $\Gamma$-point, the integral in Eq. (\ref{eq:KCNDef}) 
  over the $k$-space $\mathbb R^2$ converges stably towards the KCN of the lattice regularized model.

\begin{figure}[tp]    
    \includegraphics[width=9.cm]{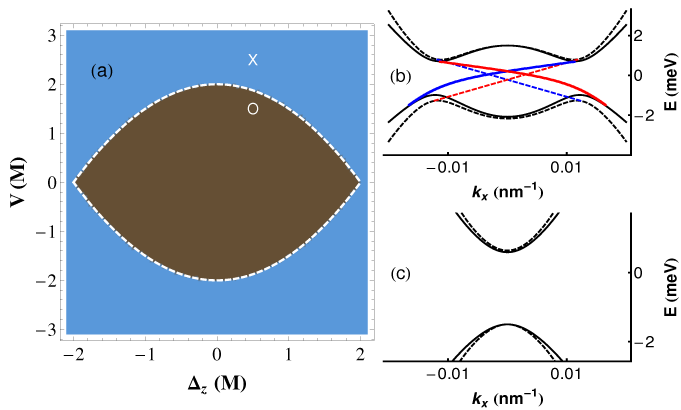}
    \caption{(Color online)
      (a) KCN of a DQW as a function of $V$ and $\Delta_z$ with $\alpha \ne 0$ and the physically plausible condition $\Delta_0=\Delta_z$, 
      corresponding to $\Delta_{H1}\approx0$ (see Appendix~\ref{sec:level2}).
      The dashed line, indicating the threshold bias is obtained with Eq.~\ref{eq:Vs}.
      The internal region is topologically trivial with $\mathcal C_\uparrow=0$ ($\mathcal C_\uparrow=2$) for $M>0$ ($M<0$), while the outer one is topologically non-trivial with $\mathcal C_\uparrow=1$.
      (b) and (c): first conduction and valence bulk bands and edge states (if present) dispersion curves obtained by numerically solving Eq.~\ref{eq:Htot} (full lines) 
      and analytically from the reduced model (dashed line) with $M=6$~meV and taking into account a spacing distance between the two QWs of $6$~nm, 
      leading to the estimated value of $\Delta_{E1} = 6$~meV, $\alpha = 5$~meV~nm (Table~\ref{table1} in the Appendix).
      The two situations correspond to the cross and the circle in (a), resp.
      For $V=15$~meV, the DQW is topologically non-trivial (b), while for $V=9$~meV, the system is trivial with no edge states (c).
    }
    \label{fig:C1}
  \end{figure}

  Let us first analyze the bulk dispersion curves described by Eq.~\ref{eq:Htot} to get an intuition for possible 
  local topological phase transitions~\cite{letopo2011} at the $\Gamma$-point, induced by band crossings.
  For $k=0$ we obtain the four eigenenergies
  \begin{eqnarray}
    E_{\pm,\eta}(0) = \eta M \pm \frac{1}{2}\sqrt{\left(\Delta_0 +\eta \Delta_z\right)^2 + V^2}, 
    \label{eq:k=0}
  \end{eqnarray}
  with $\eta=\pm 1$.
  Due to the interlayer tunneling, the $E1$ bands (centered at $E=M$) suffer an energy splitting, at the $\Gamma$-point, of $\sqrt{\Delta_{E1}^2+V^2}$, 
  while for the $H1$ bands (centered at $E=-M$) the splitting is $\sqrt{\Delta_{H1}^2+V^2}$.
  We define a local energy gap (LEG) at $k=0$ as  
  \begin{equation} 
    E_g(0)\hspace{-0.04cm} =\hspace{-0.04cm} 2 M\hspace{-0.04cm} - \hspace{-0.04cm}  \frac{{\rm sgn}(\hspace{-0.02cm} M\hspace{-0.02cm} )}{2}\hspace{-0.05cm} \left(\hspace{-0.1cm}\sqrt{\Delta_{E1}^2\hspace{-0.03cm}+\hspace{-0.03cm}V^2}\hspace{-0.04cm} +\hspace{-0.04cm}\sqrt{\Delta_{H1}^2\hspace{-0.03cm}+\hspace{-0.03cm}V^2}\right).
    \label{eq:LEG}
  \end{equation}
  When the LEG changes sign an inversion of bands with $E1$ and $H1$ character occurs at the $\Gamma$-point, accompanied by a possible topological phase transition.
  For a fixed set of parameters, the condition $E_g(0)=0$ defines (if existent) the critical value of the bias $V_c$ driving the phase transition:
  \begin{equation} 
    V_c^2 = \frac{1}{4 M^2} \left[ \left( 4 M^2 - \frac{\Delta_{0}^2 + \Delta_{z}^2}{2} \right)^2   - \frac{(\Delta_{0}^2-\Delta_{z}^2)^2}{4}\right].
    \label{eq:Vs}
  \end{equation}
  This is confirmed by the explicit calculation of the $\mathbb Z_2$-invariant $\nu$ with Eq.~\ref{eq:nu}, as shown in Fig.~\ref{fig:C1},
  where we calculated the KCN $\mathcal{C}_\uparrow$, as a function of $\Delta_z$ and $V$.
  $\mathcal{C}_\uparrow$ is insensitive to the value of $\alpha$, as long as $\alpha$ and/or $\Delta_z$ remains finite, 
  in which case an insulating gap arises.  
  Interestingly, for $|V|=|V_c|$, lowest bulk bands have at small $k$ a Dirac-like dispersion, while a tunable gap is developed for $|V|<|V_c|$.
  In the region $|V|<|V_c|$, the system supports an even number of pairs of helical edge modes $N=|\mathcal C_\uparrow|$, in particular $N=0$ for $M>0$ and $N=2$ for $M<0$. 
  In the outer region, instead, the DQW is topologically non-trivial and allows for a pair of time-reversal protected helical modes. 
  An example of the DQW spectrum in the two regimes, described by Eq.~\ref{eq:Htot}, is shown in Fig~\ref{fig:C1}(b) and (c) (full lines) 
  for a positive mass $M=6$~meV and a bias of $V=15$~meV and $9$~meV, resp. 
  Edge states are obtained by numerically solving Eq.~\ref{eq:Htot} with open boundary conditions.

\begin{figure}[tb]    
    \includegraphics[width=5.5cm]{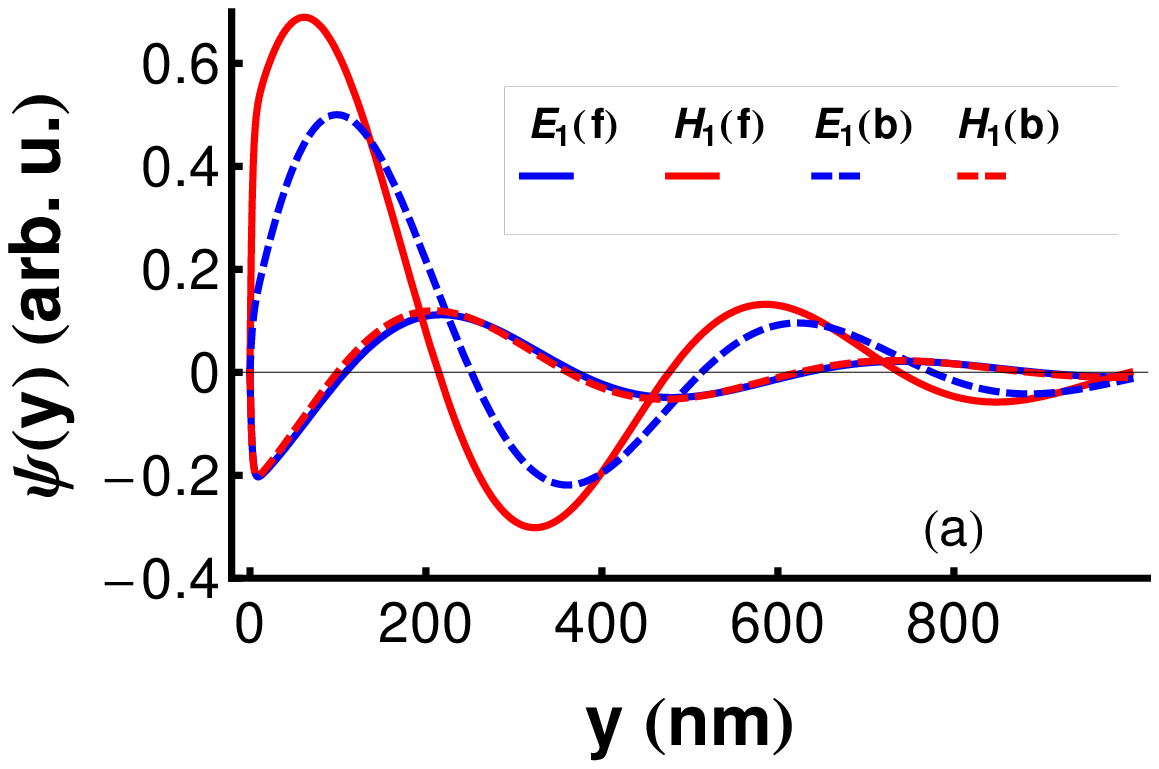}\\
    \includegraphics[width=5.5cm]{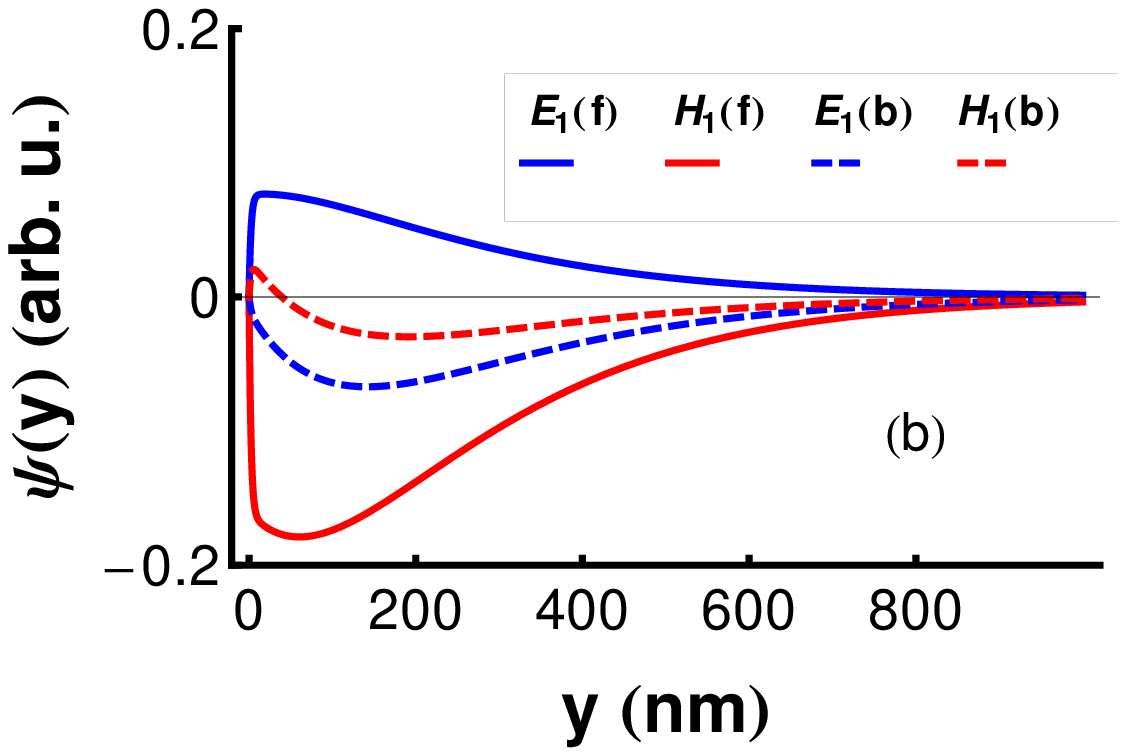}
    \caption{(Color online)
      Edge mode wave function along the Y direction, calculated from the full model with vanishing boundary condition at $y=0$, 
      for the DQW system of Fig.~\ref{fig:C1}(b) near the Dirac point $E\approx0.21$~meV in (a) and  for $E\approx-1$~meV (near the bottom of the valence bands) in (b).
      In particular, we show the behavior as a function of $y$ of the projection on the $E_1$ and $H_1$ bands of the front (f) and back (b) QWs.
    }
    \label{fig:sta}
  \end{figure}

  In Fig~\ref{fig:sta}(a) and (b),  we plot the wave function of the helical edge states for $E=0.21$~meV (near the Dirac point) 
  and $E=-1$~meV (near the valence bands), resp. for the system in Fig.~\ref{fig:C1}(b).
  These states have the peculiarity of supporting at the same time both oscillatory and decaying behavior 
  [clearly evident in Fig~\ref{fig:sta}(a)], due to the contribution of modes with complex $k_y$, which is related to the Mexican hat shape of the bulk bands.
  Approaching the bulk bands, the oscillations tend to become less pronounced, as shown in Fig~\ref{fig:sta}(b). 
  The edge modes have the major contribution from the $H_1$ band of the f layer, which consists in $60\%$ of the probability weight near the conduction band 
  and rises to $90\%$ near the valence band.
  Under the reversal of $V$, the probability weights of the b and f layer interchange.

  \section{Reduced Hamiltonian}
 In order to get an analytical understanding of the system, 
  we derive a reduced 2-band low-energy Hamiltonian $\tilde{h}_{ll}$,
  with the assumption that the relevant energies are small compared to $\frac{V}{2}\approx\pm M$.
  $\tilde{h}_{ll}$ is represented in the basis $\{|E_1,b\rangle, |H_1,f\rangle\}$ or $\{|E_1,f\rangle, |H_1,b\rangle\}$ for the $+$ and $-$ sign, resp.
  Similar to BLG, we separate the DQW Hamiltonian in Eq.~\ref{eq:Htot} in a low-energy 
  $h_{ll}$ and a high-energy $h_{hh}$ part connected by the off-diagonal blocks $h_{lh}=h_{hl}^\dag$.
  The reduced 2-band model is obtained then as $\tilde{h}_{ll} = h_{ll} +h_{lh}G_{hh}^{0}h_{hl}$, where $G_{hh}^{0}=(E-h_{hh})^{-1}$~~\cite{mccann2006}.
  The reduced model $\tilde{h}_{ll}$ can be mapped to a TI Hamiltonian of a single layer (Eq.~\ref{eq:H0}),  
  but with the renormalized parameters
  \begin{eqnarray}
    \tilde{A} &=& \frac{\alpha}{2} \pm \frac{A \Delta_z}{M \pm \frac{V}{2}} \hspace{0.7cm}
    \tilde{M} = M \mp \frac{V}{2} - \frac{\Delta_0^2 + \Delta_z^2}{4\left(M \pm \frac{V}{2}\right)} \nonumber\\
    \tilde{B} &=& B - \frac{A^2}{M \pm  \frac{V}{2}}   \hspace{0.7cm}
    \tilde{C} = C - \frac{\Delta_0 \Delta_z}{2\left(M \pm \frac{V}{2}\right)},
    \label{eq:param}
  \end{eqnarray}
  while $\tilde{D}=D$ remains unchanged.
  The bulk dispersion is characterized by a Mexican hat dispersion for $\xi \tilde{M}<\xi \frac{\tilde{A}^2}{4\tilde{B}}$, where $\xi={\rm sgn}\left[ \tilde{B}(D^2-\tilde{B}^2) \right]$,
  with a gap $E_g = -\frac{ \tilde{|A|} }{\tilde{B}^2 } \sqrt{ (\tilde{A}^2-4\tilde{M}\tilde{B} )( D^2-\tilde{B}^2 )}$.
  If the condition for the existence of the Mexican hat is not fulfilled, the dispersion has a minimum at $k=0$ and a gap of $2\tilde{M}$. 
  For a barrier thickness of $6$~nm, we obtain $E_{g}\simeq -1.6$~meV (see Table~\ref{table2} in the Appendix for details), which does not strongly depend on $M$. 
  This suggests that the parameters obtained are in the correct ballpark for experiments.

  The bulk dispersion curves are quite well reproduced by $\tilde{h}_{ll}$, as demonstrated in Figs.~\ref{fig:C1}(b) and (c) (black dashed lines).
  However, $\tilde{h}_{ll}$ predicts edge modes with purely linear dispersions 
  $E= -\frac{\tilde{D}}{\tilde{B}}\tilde{M} +s A k_x \sqrt{\frac{\tilde{B}^2-\tilde{D}^2}{\tilde{B}^2}}$, where $s=\pm 1$ denotes the two spin-blocks 
  [colored dashed lines in Figs.~\ref{fig:C1}(b)].
  Instead, edge dispersions numerically obtained from the full model [colored full lines in Figs.~\ref{fig:C1}(b)] show a marked non-linear behavior 
  when approaching the bulk bands, with an accompanying shift of the Dirac point.

  For the reduced model the KCN can be calculated analytically since $\mathcal F$~assumes the simple form \cite{bernevig2006}
  \begin{equation}
    \mathcal F_{\mu \nu}= \frac{-i}{2}\hat d\left(\partial_{\mu} \hat d\times\partial_{\nu}\hat d\right),
  \end{equation}
  where $\hat d$~is the unit vector of $(\tilde A k_x,-\tilde A k_y,\tilde M-\tilde B k^2)$. 
  The topological features of $\tilde{h}_{ll}$ are therefore analogous to those of a single layer of 2D TI.  
  However, the bias voltage enters the Dirac mass parameter $\tilde M$~thus providing an additional knob  
  to tune a system of two trivial insulators into a single-valley Dirac system $({\tilde M}=0$) and eventually into a nontrivial TI phase. 
  Since the phase transition at $\tilde M=0$~is local at the $\Gamma$-point \cite{letopo2011}
  the analytical topological predictions of the reduced model are inherited by the full model and survive 
  a lattice regularization which makes the KCN mathematically well defined.

  Finally, we note that the proposed mechanism of a voltage bias induced band inversion in a DQW is by no means restricted to HgTe based QWs. 
  The method in principle even applies to large band gap semiconductors like GaAs ($V\simeq 1$ eV) but the presumably large inversion asymmetry, 
  not captured in our model, would have to be taken into account explicitly.

 \section{Discussion and conclusion} 
 It has been proposed earlier that, in {\it single} QWs, the topological phase transition may be induced by an effective 
  applied potential~\cite{yang2008, liu2008, li2009}.
  In Refs.~\onlinecite{yang2008,li2009}, calculations show that in a single Hg$_{1-x}$Cd$_x$Te QW, a potential bias along the well thickness of the order of hundreds 
  of meV [much larger than $V$ needed in our proposal (on the order of a few meV)] induces a band inversion.
  The experimental feasibility of this approach is arduous due to the large field required.
  In Ref.~\onlinecite{liu2008}, a type-II quantum well made of InAs/GaSb/AlSb, which has an intrinsically inverted bandstructure, 
  is considered.
  The authors show that the spatial separation of E1 and H1 bands offers a key to drive the transition from the TI to the normal phase with an applied gate bias.
  Recent experiments~\cite{knez,knez2} provided the first evidences pointing towards the presence of a TI phase in these kind of structures.  
  The proposal in Ref.~\cite{liu2008} requires the growth of a type-II QW, which is a strongly asymmetric structure interfacing two specifically chosen 
  semiconducting materials with different electron affinities (InAs and GaSb).
  On the contrary, we envision a DQW with each well made from the {\it same} material 
  that might be more feasible to experimentally realize clean samples and offers the possibility of tuning the tunneling 
  parameters by modifying the barrier thickness or its chemical composition. 
  This tunability could also be used to create contacts between two helical Luttinger liquids when {\it each} QW is in the non-trivial regime and $V\approx 0$. 
  This contact can be either of electrostatic nature which allows to study Coulomb drag \cite{drag} or a new type of inter-edge correlated 
  liquid \cite{contacthelical}, or can induce the tunneling between helical edge states \cite{combiedge}.

  Further, we note that the LPS (${\vec {\cal P}}$) [see Eq.~(2)], 
  that operates on the wave function amplitude on the $(f)$ and $(b)$ layers, 
  can be manipulated by the various parameters, in particular by the voltage $V$. 
  As already mentioned, in the regime $V/2\approx\pm M$, the low-energy physics of the DQW is described by Eq.~(1), with parameters from Eq.~(10), 
  where the basis is given by eigenstates of the LPS with eigenvalues $f$ and $b$ resp.  
  %
  If ${\tilde M}$ in Eq.~(10) is zero,  ${\vec {\cal P}}$ points in-plane for small in-plane $k$-vector and has a Berry phase of $\pi$ like in graphene. 
  A finite mass ${\tilde M}$ tilts ${\vec {\cal P}}$ out-of plane and starts to localize ${\vec {\cal P}}$ in one of the layers. 
  This physics is actually reminiscent of bilayer graphene (BLG) in the presence of an interlayer voltage~\cite{mccann2006}, 
  except that the low-energy physics is governed by a LPS with Berry phase $2\pi$. 
  We expect that interesting proposals that utilize the LPS in BLG, e.g. in a LPS valve \cite{bilayervalve}, also apply here, 
  but with the important absence of a valley degeneracy that is potentially harmful for operating such devices \cite{bilayervalve}. 
  Also, the application of a voltage domain in BLG predicts valley-filtered edge states at the domain-wall \cite{domainwallbilayer} which in our case would 
  result in (Kramers) spin-filtered edge states without valley degeneracy at a mass(${\tilde M}$) domain---i.e. helical edge states. 
  Such mass domains would allow the {\it controlled} creation of helical edge states not only at the physical sample boundaries of DQW systems but also in their bulk.

  In summary, we have investigated  DQW structures using the BHZ model of HgTe QWs and a generic tunneling Hamiltonian connecting the layers.
  An interlayer potential bias on the order of the layer bandgap can drive a topological phase transition even if the individual QWs are in the normal regime.
  We calculate the $\mathbb Z_2$ topological invariant and the helical edge states which in a reduced model obtain simple analytical structures. 
  These results suggest DQWs as potential candidates for an all tunable topological insulator or Dirac system which would have desirable properties for applications.

  \begin{acknowledgements}
  We acknowledge financial support by the Emmy-Noether program of the DFG (PM and PR), from the DFG-JST Research Unit "Topological electronics" (JB) and from the DFG grant AS327/2-2 (EN). 
  We also thank H. Buhmann, L. Molenkamp and B. Trauzettel for fruitful discussions.
  \end{acknowledgements}

\appendix

\section{Estimate of the tunneling parameters in a HgTe DQW}

In the present Appendix, we will provide a quantitative calculation of the 
tunneling matrix elements [Eq.\ref{eq:tunneling0}] between the electronic states belonging to the front (f) and 
 to the back (b) HgTe/\cmtx{} quantum wells (QWs) of a double quantum well (DQW) structure.  
In the first section, we recall the $k\cdot p$ band structure calculations for the confined states of an individual HgTe/\cmtx{} QW. 
In the second section, we show how to calculate the tunneling matrix elements arising in a double quantum well configuration, 
starting from the knowledge of the envelope functions of a single QW and present the numerical values of the single-particle tunneling amplitudes.

\subsection{\label{sec:level1} Band structure model}

For the calculation of the band structure and wave functions of the single HgTe/\cmtx{} QW, an envelope-function
approximation~\cite{Burt99}, based on an eight-band $\mathitb{k}\cdot\mathitb{p}$ Hamiltonian, is used. 
The total wave function
is given as follows:
\begin{equation}\label{Psi}
  \Psi_{ \mathitb{k_\|}}(\mathitb{r})= e^{i\mathitb{k_\|}\mathitb{r_\|}}~ \sum_{n}f_{n;\mathitb{k}_\|}(z)~ u_{n}(\mathitb{r}),
\end{equation}
where $f_{n;\mathitb{k_\|}}(z)$ are the envelope functions, $\mathitb{k_\|}=(k_x,k_y)$ is the wave vector in the
plane of the QW, and $u_{n}(\mathitb{r})$ is the usual basis set for the eight-band Kane model \cite{Kane57,Kane66,Winkler03} which is assumed to be the same in HgTe- and \cmtx{}-layers:
\begin{eqnarray}
  u_{1}(\mathitb{r}) \trieq |\Gamma_6,+1/2\rangle &=&  S\uparrow \nonumber\\
  u_{2}(\mathitb{r}) \trieq |\Gamma_6,-1/2\rangle &=&  S\downarrow \nonumber\\
  u_{3}(\mathitb{r}) \trieq |\Gamma_8,+3/2\rangle &=&
(1/\sqrt{2}) (X+iY)\uparrow \nonumber\\
  u_{4}(\mathitb{r}) \trieq |\Gamma_8,+1/2\rangle &=&
(1/\sqrt{6}) [(X+iY)\downarrow -2Z\uparrow] \nonumber\\
  u_{5}(\mathitb{r}) \trieq |\Gamma_8,-1/2\rangle &=&
-(1/\sqrt{6}) [(X-iY)\uparrow   +2Z\downarrow] \nonumber\\
  u_{6}(\mathitb{r}) \trieq |\Gamma_8,-3/2\rangle &=&
-(1/\sqrt{2}) (X-iY)\downarrow \nonumber\\
  u_{7}(\mathitb{r}) \trieq |\Gamma_7,+1/2\rangle &=&
(1/\sqrt{3}) [(X+iY)\downarrow +Z\uparrow] \nonumber\\
  u_{8}(\mathitb{r}) \trieq |\Gamma_7,-1/2\rangle &=&
(1/\sqrt{3}) [(X-iY)\uparrow -Z\downarrow]. \nonumber\\
  \label{BasisSet}
\end{eqnarray}
\begin{figure} [tb]
  \includegraphics[width=5.5cm]{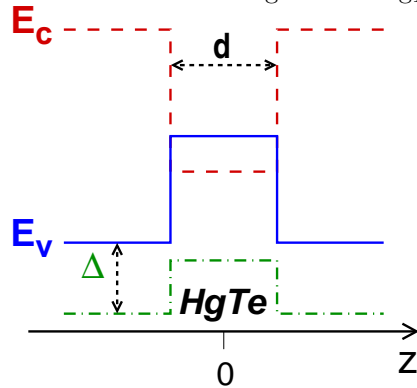}
  \caption{(Color online)
    \label{profiles}
    Band edge profile of a HgTe/\cmtx{} QW of thickness $d$. 
    The profiles of conduction and valence band are shown together with the split-off band (with the split-off parameter $\Delta$). 
  }
\end{figure}
The Hamiltonian for a QW grown along [001] direction is then given by:\cite{novik2005}
\begin{widetext}
\begin{equation}\label{Hamilt}
H =
\begin{pmatrix}
  T & 0 & -\frac{1}{\sqrt{2}}Pk_{+} &  \sqrt{\frac{2}{3}}Pk_{z} & \frac{1}{\sqrt{6}}Pk_{-} &          0                & -\frac{1}{\sqrt{3}}Pk_{z} & -\frac{1}{\sqrt{3}}Pk_{-} \\
  0 & T &        0                  & -\frac{1}{\sqrt{6}}Pk_{+} & \sqrt{\frac{2}{3}}Pk_{z} & \frac{1}{\sqrt{2}}Pk_{-}  & -\frac{1}{\sqrt{3}}Pk_{+} &  \frac{1}{\sqrt{3}}Pk_{z} \\
  -\frac{1}{\sqrt{2}}k_{-}P & 0 & U+V & -\bar{S}_{-} & R & 0 & \frac{1}{\sqrt{2}} \bar{S}_{-} & -\sqrt{2}R \\
   \sqrt{\frac{2}{3}}k_{z}P & -\frac{1}{\sqrt{6}}k_{-}P & -\bar{S}^{\dag}_{-} & U-V & C & R & \sqrt{2}V & -\sqrt{\frac{3}{2}}\tilde{S}_{-} \\
  \frac{1}{\sqrt{6}}k_{+}P & \sqrt{\frac{2}{3}}k_{z}P & R^{\dag} & C^{\dag} & U-V & \bar{S}^{\dag}_{+} & -\sqrt{\frac{3}{2}}\tilde{S}_{+} & -\sqrt{2}V \\
  0 & \frac{1}{\sqrt{2}}k_{+}P & 0 & R^{\dag} & \bar{S}_{+} & U+V & \sqrt{2}R^{\dag} & \frac{1}{\sqrt{2}}\bar{S}_{+} \\
  -\frac{1}{\sqrt{3}}k_{z}P & -\frac{1}{\sqrt{3}}k_{-}P & \frac{1}{\sqrt{2}}\bar{S}^{\dag}_{-} & \sqrt{2}V & -\sqrt{\frac{3}{2}}\tilde{S}^{\dag}_{+} & \sqrt{2}R & U-\Delta & C \\
  -\frac{1}{\sqrt{3}}k_{+}P & \frac{1}{\sqrt{3}}k_{z}P &  -\sqrt{2}R^{\dag} & -\sqrt{\frac{3}{2}}\tilde{S}^{\dag}_{-} &  -\sqrt{2}V & \frac{1}{\sqrt{2}}\bar{S}^{\dag}_{+} & C^{\dag} &
  U-\Delta
\end{pmatrix},
\end{equation}
\end{widetext}
where
\begin{eqnarray}\label{MatrEl}
  T &=& E_{c}(z)+\frac{\hbar^{2}}{2m_{0}}\left((2F+1)k_{\|}^{2}+k_{z}(2F+1)k_{z}\right),\nonumber\\
  U &=& E_{v}(z)-\frac{\hbar^{2}}{2m_{0}}\left(\gamma_{1}k_{\|}^{2}+k_{z}\gamma_{1}k_{z}\right),\nonumber\\
  V &=&         -\frac{\hbar^{2}}{2m_{0}}\left(\gamma_{2}k_{\|}^{2}-2k_{z}\gamma_{2}k_{z}\right),\nonumber\\
  R &=&         -\frac{\hbar^{2}}{2m_{0}}\left(\sqrt{3}\mu k_{+}^{2}-\sqrt{3}\bar{\gamma}k_{-}^{2}\right),\\
  \bar{S}_{\pm} &=&
  -\frac{\hbar^{2}}{2m_{0}}\sqrt{3}k_{\pm}\left(\{\gamma_{3},k_{z}\}+[\kappa,k_{z}]\right),\nonumber\\
  \tilde{S}_{\pm} &=&
  -\frac{\hbar^{2}}{2m_{0}}\sqrt{3}k_{\pm}\left(\{\gamma_{3},k_{z}\}-\frac{1}{3}[\kappa,k_{z}]\right),\nonumber\\
  C &=& \frac{\hbar^{2}}{m_{0}}k_{-}[\kappa,k_{z}],\nonumber\\
  k_{\|}^{2} &=& k_{x}^{2}+k_{y}^{2},~~~~k_{\pm}=k_{x}\pm ik_{y}\nonumber.
\end{eqnarray}
Here, the band structure parameters $\gamma_1$, $\gamma_2$, $\gamma_3$,
$\mu=\left(\gamma_3-\gamma_2\right)/2$, $\bar{\gamma}=\left(\gamma_3+\gamma_2\right)/2$, $\kappa$ and $F$ describe remote
band contributions; $P$ is the Kane momentum matrix element; $E_{c}(z)$ and $E_{v}(z)$ are the
conduction and valence band edges, respectively; $\Delta$ is the
spin-orbit splitting energy. It should be noted, that the in-plane wave vector ($k_x$,$k_y$) is a good quantum number, 
but $k_z$ should be replaced by the operator $k_{z}=-i\partial/\partial z$.

The band structure parameters for HgTe and \cmtx{} are considered as piecewise
constant for each of the layers with an abrupt change at the interfaces, according to the band edge profile in Fig.~\ref{profiles}. 
Using the correct operator ordering in the Hamiltonian [Eqs.(\ref{Hamilt}) and (\ref{MatrEl})], 
in accordance with the envelope-function approach derived by Burt\cite{Burt99},
provides us with an unambiguous determination of the interface boundary conditions. 
A detailed description of the model, as well as the values of the band 
structure parameters for HgTe and CdTe, is given in Ref.~\onlinecite{novik2005}.
\begin{figure} [b]
  \includegraphics[width=7.5cm]{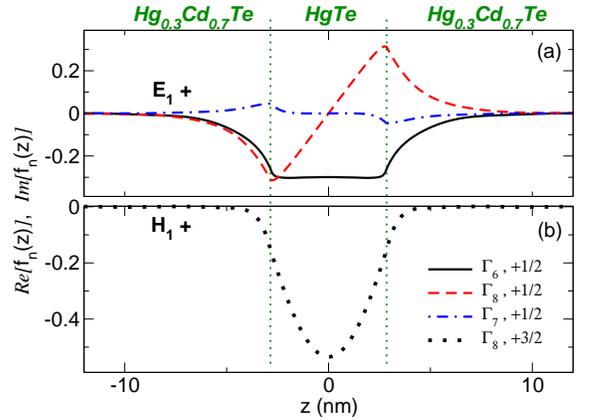}
  \caption{(Color online)
    \label{envelope}
    Envelope functions of the E1 and H1 bands of a HgTe QW of thickness $d=5.7$~nm, in panels (a) and (b), respectively, at $ \mathitb{k_\|}=0 $.
  }
\end{figure}

Solving the eigenvalue problem $H \mathbf{f}=E \mathbf{f}$ (with $\mathbf{f}$ an eight-component envelope function vector) we determine
the envelope functions $f_{n;\mathitb{k_\|}}(z)$ and energy levels near $\mathitb{k_\|}=0$ for a single QW. 
In particular, we are interested in the first conduction and valence subbands which for the QW width close to the critical value $6.3$~nm are denoted by E1 and H1~\cite{bernevig2006}. 
Other subbands are sufficiently apart, so that the system can be described, in a reasonable spectral range~\cite{Schmidt2009}, by a 4-band effective Hamiltonian~\cite{bernevig2006},  
with the basis given by the E1 and H1 subband eigenstates at $\mathitb{k_\|}=0$ ($|E1,+\rangle$, $|E1,-\rangle$, $|H1,+\rangle$, $|H1,-\rangle$).
Using the eight-band $\mathitb{k}\cdot\mathitb{p}$ model described above we calculate these states:
\begin{eqnarray}\label{Basis}
  \langle\mathitb{r} |E1,+\rangle&=&\sum_{n=1,4,7}f_{n;\mathitb{k_\|}=0}^{E1,+}(z)~ u_{n}(\mathitb{r}),\nonumber\\
  \langle\mathitb{r}|E1,-\rangle&=&\sum_{n=2,5,8}f_{n;\mathitb{k_\|}=0}^{E1,-}(z)~ u_{n}(\mathitb{r}),\\
  \langle\mathitb{r}|H1,+\rangle&=& f_{3;\mathitb{k_\|}=0}^{H1,+}(z)~ u_{3}(\mathitb{r}),\nonumber\\
  \langle\mathitb{r}|H1,-\rangle&=& f_{6;\mathitb{k_\|}=0}^{H1,-}(z)~ u_{6}(\mathitb{r}).\nonumber
\end{eqnarray}
The summation index $n$ usually runs over all eight basis states (see Eq.(\ref{BasisSet})). 
But for $\mathitb{k_\|}=0$ only envelope functions with index $n$ given in Eqs.(\ref{Basis}) have nonzero values. 
In Fig.~\ref{envelope}(a) and (b), we present the envelope functions for E1 and H1 subbands at $\mathitb{k_\|}=0$.  
Note that $f_{7;\mathitb{k_\|}=0}^{E1,+}(z)$ and $f_{8;\mathitb{k_\|}=0}^{E1,-}(z)$ give negligibly small contributions to the total wave 
function ($|\langle f_{7;\mathitb{k_\|}=0}^{E1,+}(z)|f_{7;\mathitb{k_\|}=0}^{E1,+}(z)\rangle|$, $|\langle f_{8;\mathitb{k_\|}=0}^{E1-}(z)|f_{8;\mathitb{k_\|}=0}^{E1-}(z)\rangle |<0.01$).

\subsection{\label{sec:level2} Tunneling matrix in a double quantum well}

\begin{figure} [tbp]
  \includegraphics[width=7.5cm]{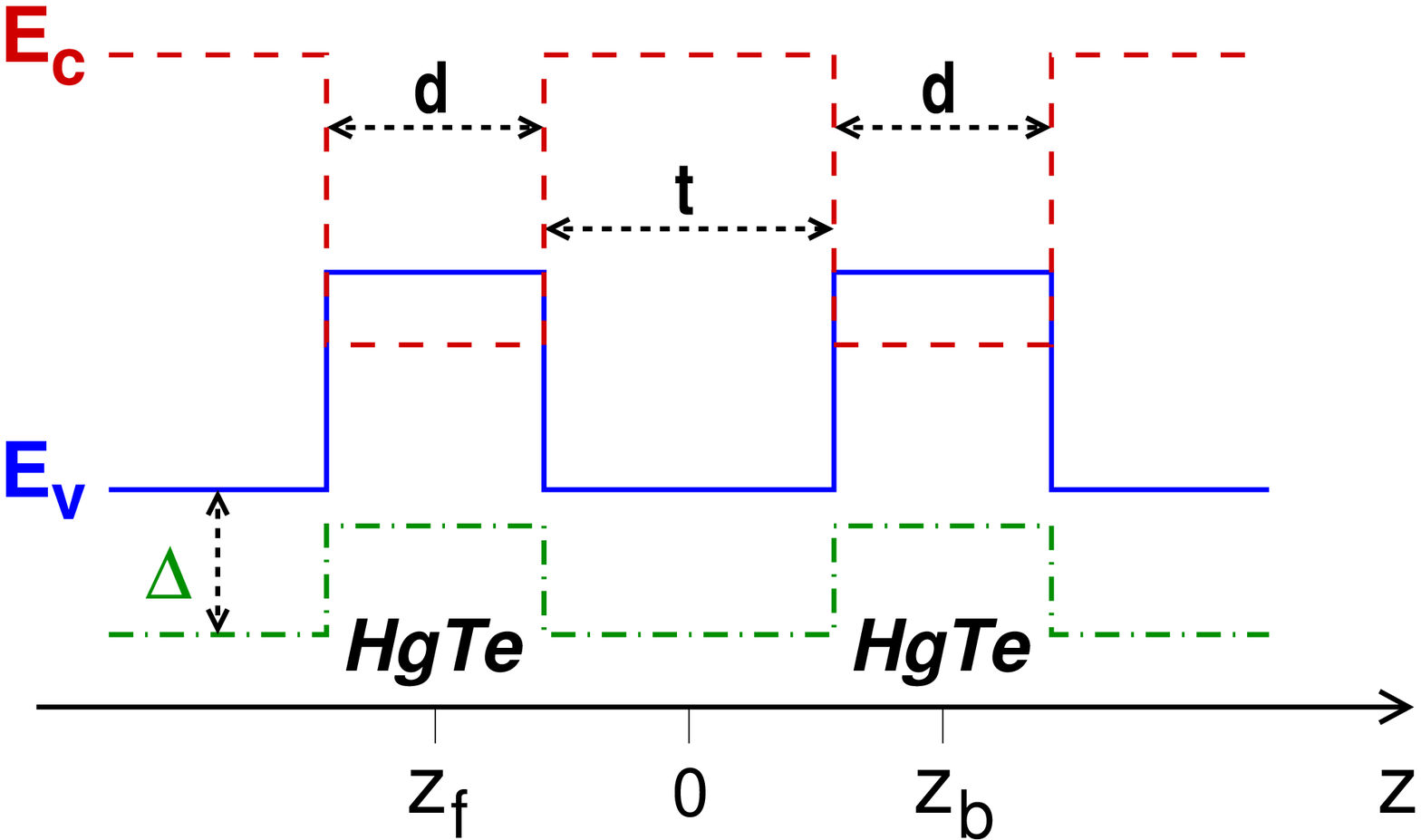}
  \caption{(Color online)
    \label{profile} 
    Sketch of the band edge profiles of a HgTe/\cmtx{} DQW. 
    The profiles of conduction and valence band are shown together with the split-off band (with the split-off parameter $\Delta$).
    $d$ is the thickness of a single HgTe QW and $t$ is the barrier thickness of \cmtx{}.
  }
\end{figure}

We have proposed an extension of the 4-band effective Hamiltonian~\cite{bernevig2006} for a HgTe/\cmtx{} QW to a DQW structure.
Here, we present a realistic estimation of the parameters in the model equation Eq.~\ref{eq:tunneling0}. 
We consider a DQW geometry as described in Fig.~\ref{profile}, where we assume a symmetric structure for simplicity. 
Two HgTe QWs of individual thickness $d$ are separated by a barrier of length $t$ of \cmtx{}.
When the barrier is sufficiently thin, the electronic states belonging to different layers share a finite overlap.
The envelope function of the DQW can be described as a bonding and antibonding combination of the envelope functions of the individual f and b QWs.
Each QW has two confined states in the spectral range of interest: the E1 and the H1 subbands.
Using the envelope functions for the E1 and the H1 bands discussed in Section~\ref{sec:level1} (see Eqs.~\ref{Basis} and Fig.~\ref{envelope}), 
we can write the tunneling matrix between the states localized on f and b wells as  
\begin{eqnarray}
  h_T &=& \langle X,+,f| \mathcal{H} |X,+,b\rangle
  \label{eq:tunneling}\\
  \tilde{h}_T &=& \langle X,+,f| \mathcal{H} |X,-,b\rangle, 
  \label{eq:tunneling_rashba}
\end{eqnarray}  
with  
\begin{eqnarray}
   \langle\mathitb{r}|X,\pm,b\rangle &=& \sum_n f_{n;\mathitb{k_\|}=0}^{X,\pm}\left(z-z_b \right)~ u_{n}(\mathitb{r}) \\
  \langle\mathitb{r} |X,\pm,f\rangle &=& \sum_n f_{n;\mathitb{k_\|}=0}^{X,\pm}\left(z-z_f \right)~ u_{n}(\mathitb{r})
\end{eqnarray}
where $z_f=-\frac{t+d}{2}$ and $z_b=\frac{t+d}{2}$ are the centers of the f and b QWs, and $X$ refers to the E1 and the H1 subbands.
$h_T$ is the tunneling matrix previously introduced in our model (in the form of $H_T = h_T \mathcal{P}_x$), 
 while $\tilde{h}_T$ is an additional Rashba-like term, mixing the two Kramer's block.
$\mathcal{H}$ is the Hamiltonian of the DQW system, which, in the eight-band Kane model basis, 
has the same form as Eq.~(\ref{Hamilt}) and follows the band edge profiles of Fig.~\ref{profile}, 
with its material-related parameters tuned to their HgTe values in the two intervals $[-d-\frac{t}{2},-\frac{t}{2}]$ and $[\frac{t}{2},d + \frac{t}{2}]$, and 
to the Hg$_{0.3}$Cd$_{0.7}$Te values elsewhere. 
Note that we neglect the distortion of the band edges due to the electrostatic potential $V$, which in our proposal contributes just a negligible correction
with respect to the band offsets between HgTe and \cmtx.
We stress that the results of any concrete calculation is obtained for a Cd content in the barriers of $x=0.7$. 
The parameter $x$ offers a way to tune the transparency of the barrier.
In particular a smaller $x$ leads to reduced band edge offsets between the QW and the barrier region (see Fig.~\ref{profile}). 

Using the envelope functions of the E1 and H1 subbands, calculated in Section~\ref{sec:level1}, we calculate the expectation values 
(integration over $z$) in Eq.(\ref{eq:tunneling}) in order to obtain $h_T$.
As expected from the band symmetry (see the related discussion in the note \onlinecite{symmetry}), $h_T$ has the following structure
\begin{equation}
  h_T=\frac{1}{2}\left(
    \begin{array}{cc}
      ~\Delta_{E1}~ &  \alpha~ k_+\\
    \alpha~ k_- &~\Delta_{H1}~
  \end{array}
  \right),
  \label{eq:Ht}
\end{equation}  
with 
\begin{eqnarray*}
  \Delta_{E1} &=& 2~\langle E1,+,f| \mathcal{H} |E1,+,b\rangle\\
  \Delta_{H1} &=& 2~\langle H1,+,f| \mathcal{H} |H1,+,b\rangle\\
  \alpha k_+ &=& 2~\langle E1,+,f| \mathcal{H} |H1,+,b\rangle\\
\end{eqnarray*}

\begin{figure} [tbp]
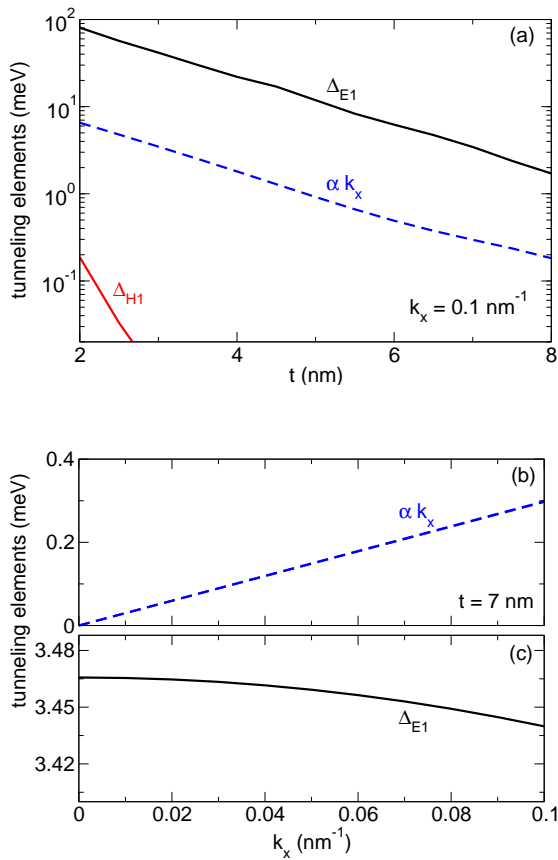

  \includegraphics[width=7.3cm]{tunneling}\\\vspace{0.8cm}
  \includegraphics[width=7.3cm]{tunnelingK}
  \caption{(Color online)
    \label{tunnel} 
    Tunneling terms of Eq.~(\ref{eq:Ht}) between the confined states of the b and f QWs of a HgTe/\cmtx{} DQW structure. 
    Tunneling terms are plotted as a function of $t$, for $k_x=0.1$~nm$^{-1}$ and $k_y=0$, in (a),  while they are shown as a function of $k_x$, for $t=7$~nm, in (b) and (c).
  }
\end{figure}

In Fig.~\ref{tunnel}(a), we show the dependence of $\Delta_{E1}$, $\Delta_{H1}$ and $\alpha k_x$ on the QW separation $t$ for a fixed value of $k_x=0.1$~nm$^{-1}$ ($k_y=0$).
As expected, both $\Delta_{E1}$ and $\alpha$ exponentially decay for increasing length of the tunneling barrier.
We note, also, that $\Delta_{H1}$ is very small and for all practical purposes it can be taken as zero. 
The quantitative difference between $\Delta_{H1}$ and $\Delta_{E1}$ is a direct consequence of the E1 band behaving as an interfacial confined state between HgTe and \cmtx{}, 
 see Fig.~\ref{envelope}.
H1 envelope function is, instead, mostly confined in the HgTe layer.

In Fig.~\ref{tunnel}(b) and (c), we show $\alpha k_x$ and $\Delta_{E1}$ as a function of the in-plane wavevector $k_x$ for $t=7$~nm. 
Parameter $\Delta_{E1}$ shows negligibly weak quadratic corrections, while $\alpha$ is strictly a constant.
This justifies the treatment of the parameters $\alpha$ and $\Delta_{E1}$ as constants in Eq.~\ref{eq:tunneling0}.
Table~\ref{table1} provides the values of $\alpha$ and $\Delta_{E1}$ for several values of  $t$, for a QW of width $d=5.7$~nm.

\begin{table}[h]
\begin{tabular}{c|ccccccccc|}
  $t$~ (nm) &~ &~~2~~ & ~~3~~ & ~~4~~ & ~~5~~ & ~~6~~ & ~~7~~ \\
 \hline
 ~$\Delta_{E1}$ (meV) &~ &80 & 42 & 22 & 12 & 6 & 3.4  \\
 ~~~$\alpha$ (meV nm)&~ &66 & 32 &  18 & 10 & 5 & 3 
\end{tabular}
\caption{Tunneling parameters $\alpha$ and $\Delta_{E1}$, calculated at $k_x=k_y=0$ for several values of the barrier thickness $t$. $d=5.7$~nm, corresponding to a Dirac mass term $M\approx6.5$~meV.}
\label{table1}
\end{table}
Here should be noted that BHZ model is applicable in a finite spectral range~\cite{Schmidt2009}, this fact restricts the maximum value of the tunneling matrix elements
for which the approach we follow is valid.
We performed calculations for $d$ in a range between $5.7$~nm and $6.6$~nm, observing that the tunneling elements smoothly vary. 
In particular, for increasing $d$, $\Delta_{E1}$ is slightly reduced (there is a variation of about $10\%$ between values corresponding to QW widths of  $5.7$~nm and $6.6$~nm).
Parameters $\alpha$ and $\Delta_{H1}$ are sensibly increasing with $d$, however in all the explored range $\Delta_{E1}>>\Delta_{H1}$.

\begin{figure} [tbp]
  \includegraphics[width=7.5cm]{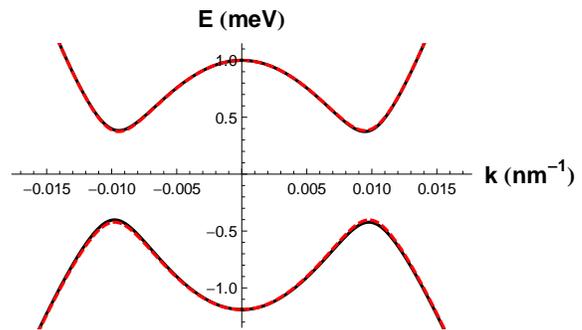}
  \caption{(Color online)
    \label{disp} 
    Particular of the fist conduction and first valence bands of a DQW structure with $t=7$~nm, $d=5.7$~nm, obtained from our model, 
    and including the Rashba-like tunneling term in Eq.~(\ref{eq:tunneling_rashba}). 
    Full lines and dashed lines refer to the two Kramer's blocks.
  }
\end{figure}

The DQW geometry, breaking the single QW mirror symmetry, generates also Rashba-like tunneling terms described by Eq.(\ref{eq:tunneling_rashba}), 
which are not included in our model for clarity of the presentation.
Our calculation shows that the only significant Rashba-like tunneling element is $\langle E1, +,f| \mathcal{H} |E1, -, b \rangle = \frac{\tilde{\alpha}}{2} k_-$.
However, in our proposal there is an energy potential shift $V\approx \pm 2M$ between f and b QWs.
This energy detuning between the E1 confined levels in the f and b QWs strongly hinders their coupling by the Rashba-like interaction.
In practice, when we include the Rashba-like term in the model, using the realistic tunneling elements we have calculated, 
it results in a small (compared to the bandgap) tilting of the bulk dispersion curves (see Fig.~\ref{disp}).
Due to this, the dispersion curves are characterized by $E_{\pm}(\mathitb{k_\|}) \ne E_{\pm}(-\mathitb{k_\|})$, 
with only the time reversal symmetry requirement $E_{+}(\mathitb{k_\|})=E_{-}(-\mathitb{k_\|})$ still holding.
We note that by turning off adiabatically the Rashba-like term, we can connect the system with Rashba-like interactions 
to that previously presented in the main body of this work without closing the bandgap.
This proves that the topological properties of the system are not affected by the Rashba-like tunneling term in Eq.~(\ref{eq:tunneling_rashba}).    
\begin{table}[tb]
\begin{tabular}{c|cccc}
  $t$~ (nm) &~~ & ~~~5~~~ & ~~~6~~~ & ~~~7~~~ \\
 \hline
 ~$E_g$ (meV) &~~ & -3.9 & -1.6 & -0.85 \\ 
\end{tabular}
\caption{Bulk gap originated from the tunneling elements in the DQW structure with individual QW width $d=5.7$~nm, corresponding to a Dirac mass term $M\approx6.5$~meV.}
\label{table2}
\end{table}

In Fig.~\ref{disp}, we show the bulk dispersion curve of a DQW structure employing the tunneling parameters of $t=7$~nm (see Table I), 
with $M\approx6.5$~meV, corresponding to $d= 5.7$~nm, 
and with a potential shift $V=-15$~meV (we note that the electric field employed is about $1$~mV/nm, 
which should be easily achievable in semiconductor heterostructures).
Full lines and dashed lines refer to the two Kramer's blocks.
The Rashba-like term in Eq.~(\ref{eq:tunneling_rashba}) is included, with $\tilde{\alpha}\approx 6$~meV~nm, but its effect, as previously explained, 
is negligible and consists in a slight tilting of the bands.   

A semiconducting gap of $0.85$~meV arises, in agreement with the value obtained with the reduced model Eq~\ref{eq:param}, using the formula
\begin{equation}
\label{Eg}
E_g=-\frac{|\tilde{A}|}{\tilde{B}^2}\sqrt{(\tilde{A}^2-4\tilde{M}\tilde{B})(D^2-\tilde{B}^2)}.
\end{equation}
Table~\ref{table2} shows the value of the bulk gap for several values of $t$.

For $t<5$~nm, the reduced model (and therefore Eq.~\ref{Eg}) is no longer valid due to the large tunneling terms ($H_T$) compared to the individual QW mass term $M$, 
however the full model [Eq.(3)] still holds (as long as the BHZ model is a reasonable approximation of the band structure).



\begin{thebibliography}{24}
\bibitem{pauli}{To make the notation more compact, we use four component Pauli matrices  $\vec{\sigma}=\left(\mathbb{I}, \sigma_x, \sigma_y, \sigma_z \right)$.}
\bibitem{note2}{For definiteness, 
    we consider $\vec{\Delta}$ to be real in the remaining of the paper.}
\end{thebibliography}

\end{document}